\documentclass[lnbip]{svmultln}

\usepackage{makeidx}  % allows for indexgeneration
\usepackage{graphicx}
\usepackage{graphics}
\usepackage{amsmath}

\begin{document}

\mainmatter              % start of the contribution

\title{Scalable Architecture for Personalized Healthcare Service Recommendation using Big Data Lake}

\titlerunning{Scalable Architecture for Personalized Healthcare Service Recommendation}  % abbreviated title (for running head)

\author{Sarathkumar Rangarajan\inst{1} \and Huai Liu\inst{1} \and Hua Wang\inst{1} \and Chuan-Long Wang\inst{2}}

\authorrunning{S. Rangarajan, H. Liu, H. Wang and C. Wang}   % abbreviated author list (for running head)

\institute{Centre for Applied Informatics, Victoria University, Melbourne, Australia \\
\and Taiyuan Normal University, Shanxi Province, China\\
\email{sarathkumar.rangarajan@live.vu.edu.au, {Huai.Liu, Hua.Wang}@vu.edu.au,clwang218@126.com}\\
}

\maketitle            

\begin{abstract}        % give a summary of your paper
The personalized health care service utilizes the relational patient data and big data analytics to tailor the medication recommendations. However, most of the health care data are in unstructured form and it consumes a lot of time and effort to pull them into relational form. This study proposes a novel data lake architecture to reduce the data ingestion time and improve the precision of healthcare analytics. It also removes the data silos and enhances the analytics by allowing the connectivity to the third-party data providers (such as clinical lab results, chemist, insurance company,etc.). The data lake architecture uses the Hadoop Distributed File System (HDFS) to provide the storage for both structured and unstructured data. This study uses K-means clustering algorithm to find the patient clusters with similar health conditions. Subsequently, it employs a support vector machine to find the most successful healthcare recommendations for the each cluster. Our experiment results demonstrate the ability of data lake to reduce the time for ingesting data from various data vendors regardless of its format. Moreover, it is evident that the data lake poses the potential to generate clusters of patients more precisely than the existing approaches. It is obvious that the data lake provides an unified storage location for the data in its native format. It can also improve the personalized healthcare medication recommendations by removing the data silos. 
%                         please supply keywords within your abstract
\end{abstract}
\section{Introduction}
In the last two decades, our living and working environments were greatly enriched by affordable smart and mobile devices, and digital services\cite{wang2017special}. The interactions with digital services and devices will generate a huge amount of data\cite{wang2015special}. Undoubtedly, the enormous growth in the amount of data collected and stored by organization’s around the world over the past few decades is irrefutable. The ability to access and analyze this data is rapidly becoming more and more important. On the other hand, the workflow for acquiring and analyzing the data and subsequently transferring them into actionable knowledge is complex. 

\subsection{Personalized Healthcare}
Based on the recent research observations, the patients with the same diagnosis may respond to the same medication in different ways. A drug can be highly effective for one patient, whereas the same drug might not produce the expected results when given to another patient with the same diagnosis. Personalized medication means the prescription of precise treatments and therapeutics well suited for an individual taking into consideration of all the data that influence response to therapy\cite{jain2009textbook}. Due to the enormous growth of Internet of Things, healthcare industry is equiped with the smart devices and applications. Consequently, the digitalization creates valuable data about the patients and medications, namely Electronic Health Records (EHR)\cite{zhang2017health}. EHR's availability in large scale allows the researchers to unearth the possibilities to move the healthcare organizations towards the personalized healthcare\cite{wang2016detection}. 

EHR consists of not only the structured data but it contains the semi and unstructured data as well. Only 20\% of data is in a structured format which can be easily utilized by data scientist; but semi and unstructured production rate is 15 times higher than the structured one\cite{feldman2012big}. However, health IT research demands to process all kind of structured, semi and unstructured data to find the valuable insights\cite{chen2014big}. Inevitably, an improved data management system would help the data scientists to provide the tailored medication. The contemporary IT infrastructure provides many data handling systems such as Enterprise Data Warehouse (EDW)\cite{inmon2010dw}; but there is a lack of scalability because the EDW data management system is for well-known queries and clearly defined policies\cite{devlin1996data}. 

To pull the data into EDW for further processing, it should be gone through the procedure of data preprocessing namely Extract, Transform, Load (ETL)\cite{simitisis2007data}. ETL process predominantly consumes notable cost and time. To provide custom-made medical intervention, data from diverse sources need to be processed\cite{amine2016efficiency,simitsis2005extraction}. However, EDW system is not so capable of handling the various sourced data. Another contention in healthcare analytics to adapt EDW is it's inability to coexist with the contemporary programming based query languages. If an EDW once designed properly for processing certain business rules then it is too difficult to redesign for the future needs. This paper proposes a novel method for the healthcare data architecture using data lake technology as an alternative data architecture.

\subsection{Data Lake}

An emerging concept that has gained increasing popularity is the data lake. A data lake uses a flat architecture to store data in their raw format\cite{inmon2016data}. Each data entity in the lake is associated with a unique identifier and a set of extended metadata. The consumers can use purpose-built schemas for query-relevant data, which will result in a smaller set of data that can be analysed to help answer a consumer's question. There are doubts and concerns about the possibility of data becoming incomprehensible due to a lack of schema or similar means of interpretation, and that could cause the lake to turn into a "data swamp" \cite{Hai:2016:CID:2882903.2899389}. Therefore, a metadata repository that registers high-level information about data entities (type, time, creator etc.) is an essential component of a robust data lake structure. A data lake's flat structure stores data regardless of its format and places the responsibility for understanding the data elsewhere. 
	 
We are thus motivated to choose an alternative data architecture for the healthcare industry. Personal data lake system proposed by Walker and Alrehamy~\cite{walker2015personal}gives a basis for designing our proposed approach. The promising properties of personal data lake architecture are as follows:

\begin{itemize}
	\item It can provide a unified location for all the data from the different social network about a single user. 
	\item It can improve the privacy and security by storing in a single location and the user is given the rights to design how to access their data. 
	\item It also offers the Personal Lake Serialization Format (PLSF) approach for storing meta-data. 
\end{itemize}

However, the original personal data lake proposed is restricted with structured and semi-structured data and it is not able to manage the unstructured data ~\cite{walker2015personal}.

\subsection{Role of Data Lake in Healthcare}
This study proposes to adopt data lake architecture as a replacement for the traditional data management architecture in healthcare. Data lake possesses the following capabilities to address the aims of this research:

\begin{itemize}
	\item It can store the data in its native format (structured / unstructured) as arrived without any pre-processing delay.
	\item Data lake can connect with trusted external sources (clinical lab, genomic centre, insurance payers, and social media)\cite{vernon2015information}. This will reduce the data silo across health care institution \cite{henry2015big}.
	\item It can support new types of data processing and improve the adaptability of the analytics system. It can store huge amount of data from a diverse source with less cost.
\end{itemize}

\subsection{Contribution of the research}
Motivated by the above-mentioned needs and possibilities, we propose a scalable architecture for personalized healthcare recommendation. The main contributions of this study are:
\begin{itemize}
	\item It introduces the data lake architecture in healthcare to crawl and ingest healthcare data from vendors without any data preprocessing delay.
	\item It enhances the data IT infrastructure in healthcare by accepting the connection from trusted third party data stakeholders.
	\item It accumulates the data with different formats and store it in the unified data lake to avoid the Data silos across the healthcare organizations.
\end{itemize}

\section{Proposed Data Lake Architecture}
\begin{figure}
	\centering
	\includegraphics[width=0.8\textwidth]{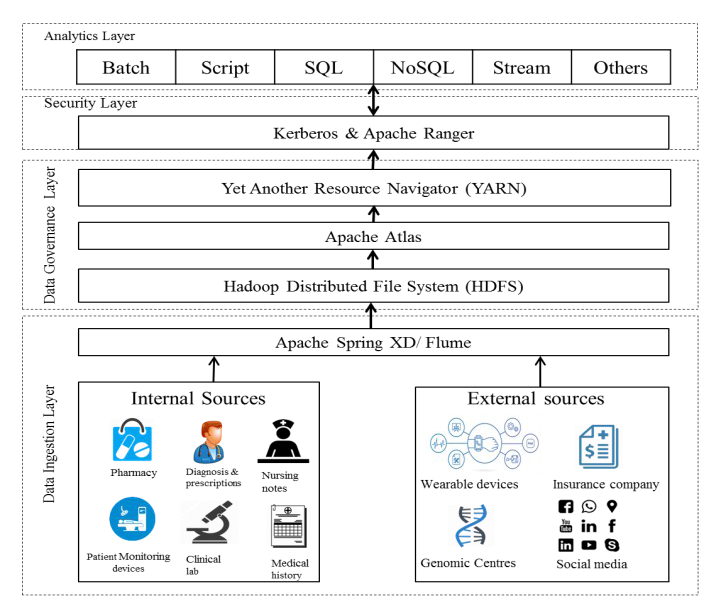}
	\caption{Data lake architecture}
	\label{fig:archi}
\end{figure}

The proposed data lake architecture for this research is plotted in Fig.~\ref{fig:archi}. It has four layers, namely, data ingestion layer, data governance layer, security layer, analytics layer, which will be respectively discussed in the following four sections.

\subsection{Data ingestion layer}
Typically, data will be crawled from multitude sources with its raw format. More often, the data will be available as structured but sometimes it may also arrive as semi-structured and unstructured. Data lake can ingest all the available data without any ETL processing but it also needs a worthy Metadata management. Because data lake without a proper metadata management will make it as a data swamp. Metadata contains information about how, when and by whom it was collected, created, accessed, modified and how it is formatted \cite{mathew2016big}. Metadata can be categorised into technical, operational, and business metadata.

Data lake can be implemented using open source software named Apache Hadoop. It has a Hadoop Distributed File System (HDFS), which allows us to store structured, semi-structured and unstructured types of files. HDFS can store petabytes of data and can act as a single storage location. It is fault tolerant, scalable, and extremely simple to expand \cite{feldman2012big}. There are many tools available to ingest data to the HDFS from sources. In the healthcare industry, three basic types of data sources are available. The first type is the bulk size of data providers, such as genomic centres and biobanks. Next type of data sources is event-based data source, such as doctor appointment, pharmacy purchase, and clinical notes. The last data source type is trusted third party streaming data\cite{li2017multi} providers such as clinical labs, X-ray centres, social media, wearable devices.

In our architecture, we have the following particular settings.
\begin{itemize}
	\item Hadoop's Spring XD tool is used to transfer bulk data\cite{kamal2017real}. It also can create metadata information while the data ingested and loaded with the HDFS.
	\item Apache Flume is a distributed, reliable, and available service for efficiently collecting, aggregating, and moving large amounts of log data\cite{begum2016rectify}. It has a simple and flexible architecture based on the data flows and it is robust and fault tolerant.
	\item To handle the stream of data and avoid data silos, we planned to utilise Hadoop spring XD tool. It can do data ingestion along with metadata information creation from multiple input sources into HDFS with high throughput.
\end{itemize}

\subsection{Data governance layer}

The main purpose of this layer to understand, organise, manage and provide access to all the data collected. This layer uses Apache Atlas, a tool for Hadoop to handle metadata framework and governance\cite{begum2016rectify}. It has a set of basic governance services to meet the compatibility requirements for the HDFS. Atlas tool has four important features, namely data classification, centralised auditing, search and lineage, and security and policy engine.
		
\subsubsection{Data Classification}
It imports or defines data-oriented metadata from the data source state, interpret and understand the relationships between data sets and core elements including source, target, and ingestion processes.
\subsubsection{Centralized Auditing}
It makes a log registry for interaction with the data stored in HDFS for reporting.
\subsubsection{Search and Lineage}
It develops a well-defined path for data exploration by recording the information about the creation of data and metadata.
\subsubsection{Security and Policy Engine}
\begin{figure}
	\centering
	\includegraphics[width=100mm]{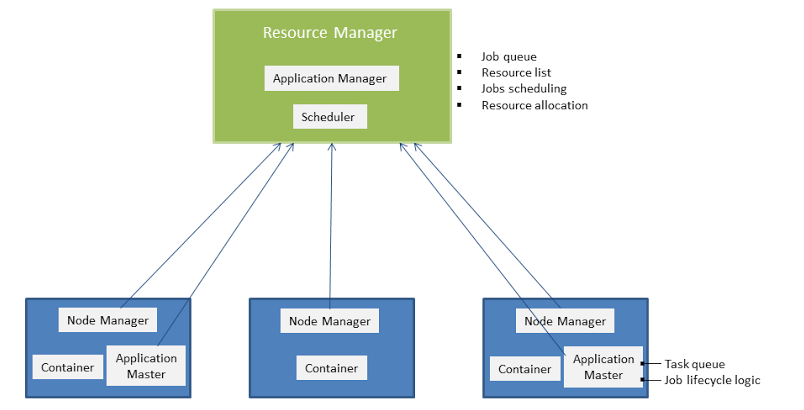}
	\caption{Work flow of YARN}
	\label{fig:workflow}
\end{figure}

It defines and justifies the data access by role based access and protects data from tampering \cite{abbas2016personalized}.
Data governance layer is also responsible for resource management and job scheduling. Data available in the data lake are not in the similar structure. Therefore, an efficient resource management is essential to make the negotiation of the data analytical programs easier. The existing healthcare analytics system uses Map Reduce methodology to schedule the CPU cycle and memory across the clusters in Hadoop to process the job\cite{archenaa2015survey}. However, MapReduce is not dealing with the scheduling of Data resource for the waiting jobs. Therefore, we utilise Apache Hadoop YARN (Yet Another Resource Negotiator) instead of MapReduce as a data operating system to let the data processing systems to interact with the data efficiently\cite{shaikh2016yarn}. Unlike MapReduce, YARN handles data processing efficiently by splitting resource management and job scheduling into separate process \cite{patel2015big}. The work flow of YARN is depicted in Fig.~\ref{fig:workflow}. 

Application Master (AM), Node Manager (NM), Resource Manager (RM) are the major actors of YARN. These actors work as follows:

\begin{itemize}
	\item Once a data processing engine request reached YARN, it allocates required resources to start AM. After starting up, AM will log its entry in RM.
	\item If any additional resource needed, then AM can negotiate with RM. Once the resource made available, AM will make the NM allocate the resources to the process.
	\item A vacuum created for the resources and the code to run within it. AM will get the execution status and it will be reported to RM too.
	\item The client can communicate with either AM or RM to get the status update. Once the code executed, AM for the respective job will remove itself from the RM and release all the allocated resources.
\end{itemize}

\subsection{Security Layer}

Obviously, the healthcare data needs highly secured environment because it contains information that is very sensitive\cite{sun2012purpose,li2008current}. Therefore, we plan to provide a more efficient security system for HDFS. Authentication and authorization are the two important processes for providing controlled access in HDFS\cite{sun2012semantic,wang2005flexible}. 
\begin{itemize}
	\item To provide authentication, we are going use Kerberos authentication protocol. Kerberos protocol will create a proxy server to receive a client request~\cite{valliyappan2016hap}. If the request is legitimate, then it will provide a ticket for the client with a timestamp.
	\item Apache Ranger is an efficient tool to provide the authorization\cite{shaw2016hive}. It is a unified authorization model for HDFS. It enables the data lake leaders to create security policies and role-based access control for the data. 
\end{itemize}

Whenever a client with a legitimate ticket enters into the system, Apache Ranger validates the ticket using its security policies. It has a very flexible user interface and is easy to deploy security policies for the huge amount of data storage.

\subsection{Analytics layer}

While comparing with Data warehouse, Data Lake is very effective at utilizing the vast amount of data with data analytical algorithms to identify the valuable insights that will improve real-time decision analytics. Since the HDFS can be connected with many number of data analytics tools, Data Lake can be adapted for the future. To evaluate our system, we planned to make the clusters of patients with similar health conditions and drug acceptance based on their detail available in the EHR and other data sources. In particular,

\begin{itemize}
	\item K-means clustering algorithm is used to do the clustering. K-means algorithm will be implemented using Matlab programming environment.
	\item The next step is to find the best medication practice for each cluster using Support Vector Machine (SVM).
\end{itemize} 

SVM is a supervised machine-learning model associated with a learning algorithm\cite{weston2014support}. Ideal medication training data to be used to train the SVM. Then the cluster will be processed by SVM and best efficient medication recommendation will be identified. Some sample data will be given to the SVM to validate its accuracy level. Once the SVM reaches 90\% of accuracy then it will act as a recommender system for the future medication recommendations.

\section{Experiments}	

We evaluated our methods via experiments on sample EMRs data. In this section, we first give a brief description of the research questions, and then explain the experimental setup to validate the usability of the proposed architecture.

\subsection{Research Questions}
We designed our experimental study to answer the following two research questions:

\subsubsection{RQ1: Can the proposed architecture reduce the time for data ingesting and crawling from various internal and external data stakeholders?}
Compared with the traditional data warehouse, the data lake allows the storage of data as it comes without bounding with any schema. In addition, it uses HDFS file system, which can connect to any remote application using Apache tools. The time taken for the data architecture to load and store the data is represented as Data ingestion time. Apparently, the less the data ingestion time the better the data architecture for healthcare analytics. Therefore, if the proposed data lake architecture can reduce the data analytics processing time, it will also improve the healthcare recommendations. 

\subsubsection{RQ2: Can the data lake architecture able to avoid Data silos?}
Due to the pre-defined data schema for data warehouse architecture, it is impossible to store the data of different types in a unified storage location. It leads to the creation of numerous data silos for the dataset about each patient. The precision of clustering is dependent on the perimeter of the data about the patients. The more data from various data stakeholders will help the data analytics to provide better results. Hence, if our proposed data lake architecture can handle the various data types in a unified location without the data swamp threat, it can improve the precision of the clustering significantly.

\subsection{Variables and objects}

\subsubsection{Independent variables} \hfill\\

Independent variable in the experiment is the technique under investigation. By nature, the proposed data lake architecture was selected for this variable. We focused on the patient clustering methodology in the architecture, as it is the vital process to identify the more suitable healthcare practice for the given patient pool. In addition, we selected the traditional data warehouse (DW) as the baseline technique for the evaluation and comparison of clustering precision. 

\subsubsection{Dependent Variables}

\paragraph{\textbf{Data Ingestion time:}}
We used data ingestion time as a metric to validate the RQ1. The ingestion time is the time taken for the data architecture to load the data to its storage. The meta data log updated by Apache spring XD each time whenever there is a new data was entered in the data architecture. The timer started when the data reaches the data architecture and it stops when the data entry created in the meta-data log. Data ingestion time IT can be calculated by finding the difference between the meta log time and the arrival time. Formula (\ref{eq1}) shows the calculation of data ingestion time for data $ k $ by subtracting the data arrival time of $ k $ from the meta data log entry time of $ k $.

\begin{equation}\label{eq1}
IT_k=MLtime_k-DAtime_k,
\end{equation}
where $IT_k$ refers to the data ingestion time of $k$th data, $MLtime_k$	represents the data log entry time for data $k$, and $DAtime_k$	is the data arrival time of data $k$

\paragraph{\textbf{Clustering precision:}}
In pattern recognition studies, the importance goes to finding out the relevance between patterns falling in n-dimensional pattern space. To find out the relevance between the patterns, it is important to examine the characteristic distance between them. The characteristics distance between the patterns decides the unsupervised classification (clustering) criterion. As per the theory, we considered the Euclidean distance to evaluate the precision of the clustering\cite{ghosh2013comparative}.

\begin{equation}\label{eq2}
{\text{Euclidian distance }d=\sqrt{\sum_{i=1}^{n}(x_i-y_i)^2}}
\end{equation}

We used the Euclidian distance as a metric to for RQ2. For the same collection of patient records, the patients clustering need to created. The distance between two points can be calculated using Formula (\ref{eq2}), where, $x_i$ and $y_i$ are the $i$th coordinates for points $x$ and $y$, respectively, and $d$ is the distance between $x$ and $y$.

The clustering metric $d$ was calculated for all the available clusters created by both DW architecture and proposed data lake architecture. The lower value of $d$ indicates the higher precision of clustering. Obviously, the higher cluster precision implies a more effective data architecture for the healthcare recommendation system.

\subsection{Objects}

We made use of anonymized EHR~\cite{sun2012satisfying} and its supporting data to evaluate our data lake architecture. We made use of UCI machine learning repository \cite{strack2014impact}, which contains the data set of diabetes form 130 US hospitals during the years of 1999-2008. It has 10 years of inpatient encounters from 130 US hospitals and integrated delivery networks. It contains 50 features representing patient and hospital outcomes. The data contains such attributes as patient number, race, gender, age, admission type, time in hospital, medical specialty of admitting physician, number of lab test performed, HbA1c test result, diagnosis, number of medications, diabetic medications, as well as the number of outpatients, inpatients, and emergency visits in the year before the hospitalization, etc. 

\subsection{Empirical environment}

The data available in the data set was classified as internal sourced data and external sourced data. The internal source data was connected with HDFS system Apache spring XD and loaded into the data lake. The external sourced data was connected with data lake system by using Apache Flume. Apache atlas identified the metadata available with data from the source. Each data would be allocated with a unique identifier to easy access. K-means algorithm using Matlab was performed on the available data to identify the available clusters. The identified cluster would contain the patients with similar health conditions. SVM trained with the training data from the data set. SVM run on each cluster to find the most successful medication recommendation. A new data would be given to the system after the authorization. The personalized recommendation for the new data have been identified from the SVM.

\section{Experimental Results}

This section describes the performance of proposed data lake architecture compared with the data warehouse.

\subsection{Reduction of Data Ingestion Time}

\begin{figure}
	\centering
	\includegraphics[width=0.50\textwidth]{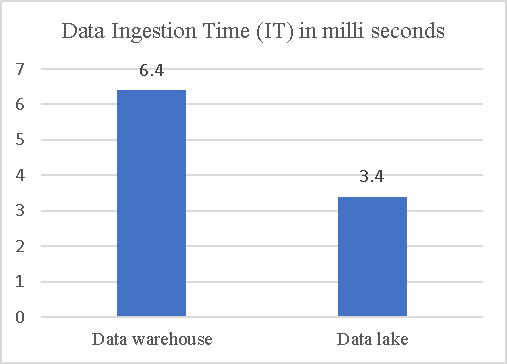}
	\caption{Comparison of data ingestion time of DW and data lake}
	\label{fig:graph}
\end{figure}

 We collected the data arrival time and Meta data log entry time for all the data tuples from the dataset in both DW and our data lake architecture. We then calculated the data ingestion time based on Formula (1). The average values of the ingestion time for DW and the data lake architecture are plotted in Fig.~\ref{fig:graph}. It is clearly shown that the data lake architecture has much lower average value of IT.

\subsubsection{Answer to RQ1:} 
The experimental results clearly show the proposed approach's ability to store the data even from those in the native form. The data ingestion time has been improved significantly by the data lake architecture. Since the proposed architecture makes use of the HDFS file system, it does not require the preprocessing stage for the data. By contrast, the DW technique involves the ETL process, which takes much more time to ingest the data into the data warehouse.

The Apache Flume has the ability to pull the data from the remote data vendors and to store successfully in the data lake environment. The Kerberos engine creates an authentication ticket for each login, and the Apache Ranger tool verifies the authentication tickets and then provides the access rights to the remote login. These steps enable the third party data stakeholders to connect with the data lake architecture with security.

\subsection{Removal of Data Silos}
We compared the clustering algorithm based on the precision quality. We identified four clusters with maximum data points for each data architecture. The data points (namely, $x$ and $y$) were further identified for the calculation of the precision value. In particular, the Euclidean distance d between $x$ and $y$ was calculated according to Formula (2). Table 1 summarises the values of d for each cluster. The results clearly show that the data lake architecture has smaller values of d for clustering than DW. In other words, the proposed data lake architecture has higher precision in the clustering.

\subsubsection{Answer to RQ2:}
The precision of clustering normally increases as the amount of data becomes larger. However, due to its schema and organizational rules of DW, much vital information about patients will be lost during the ETL process. By contrast, the data lake architecture makes use of the schema-on-read method to load the data into the environment, which bring in the ability to connect with the third-party and external data stakeholders. The high availability of data about patients helps improve the precision of clustering. In a word, the data lake architecture delivers the ability to store a variety of data within the unified location. Thus, the clustering precision has been significantly improved.

\begin{table}[]
	\centering
	\caption{Comparison of precision value $d$ for DW and data lake}
	\label{precision table}
	\begin{tabular}{|c|c|c|}
		\hline
		\multicolumn{1}{|l|}{\textbf{Euclidean distance}} & \multicolumn{1}{l|}{\textbf{$d$ value of data lake}} & \multicolumn{1}{l|}{\textbf{$d$ value of DW}} \\ \hline
		Cluster 1                                         & 0.64                                               & 0.76                                                    \\ \hline
		Cluster 2                                         & 0.71                                               & 0.87                                                    \\ \hline
		Cluster 3                                         & 0.65                                               & 0.87                                                    \\ \hline
		Cluster 4                                         & 0.88                                               & 0.99                                                    \\ \hline
	\end{tabular}
\end{table}

\section{Related work}
Undoubtedly, we are living in the world of smart devices, which create data for almost everything. Reuters predicted the global growth of data in 2020 will be nearly 35 Zettabytes (one Zettabyte = one million petabytes) through all the electronified organizations. It is three times higher than the amount of data we produced in 2015. 

Due to this enormous usage of data all over the world, the term “Big Data” is widespread and starting to be a part of every business process. Collecting and storing this vast scale of data is worth nothing without retrieving useful knowledge through processing and analysing it \cite{henry2015big}.Unlike the traditional data generation models, big data contains a variety of data from the diverse sources with high velocity in huge volumes.

The evolution of Big Data created the conditions for transforming the Healthcare industry towards the Electronic Health Records (EHR) and managing computerized archives. The EHR may contain the information about various attributes of the patients, which could include demographic details, previous medication history, and allergies, vaccination status, laboratory test results, EMR scan reports, payer information, insurer details, a previous visit to hospitals and so on \cite{katehakis2006electronic}. 

Therefore, it is obvious that the research on EHR and patient related auxiliary data will illuminate the improvised healthcare at the point of care. Inevitably, big data analytics on healthcare will enhance clinical operation by providing more relevant, efficient, error-free and cost effective diagnosis and medication. Since the availability of the data also shared with the patient, it will increase the transparency and reliability of the medical institutions \cite{feldman2012big}.

The proactive, patient-centric and tailored healthcare medication recommendation to the individual based on the analysis of the corresponding person’s EHR, genomic profile, laboratory data and other related supporting data \cite{yoon2017discovery}. 

The healthcare organisations create structured, semi-structured and unstructured data in the form of EHR. Even more importantly, only 20\% data is in a structured format, which can be easily utilised by data scientist using the analytics machines but semi / unstructured production rate is 15 times higher than the structured one \cite{feldman2012big}. 

Naturally, the sources for the data also range from mobile devices, social media feeds wearable devices, Radio Frequency Identification (RFID) devices, sensors and monitoring devices attached to the patient and their bed \cite{mathew2016big}. Even though the health IT has seen tremendous technology development, it is challenging to manage this complex data flood.

Systems like CARE (Collaborative Assessment and Recommendation Engine), ICARE predicts the future disease risk of the patient by analysing patient's previous history \cite{davis2008predicting}. HealthCare ND is also a system to predict the future disease risk of the patient \cite{dentino2010healthcarend}. However, it is an interactive system, which will get the health-related information from the patient and process the data with ICD-9-CM codes then return the results of the patient's future prediction. Abbas et al. proposed a system called Collaborative Filtering based Disease Risk Assessment (CFDRA) \cite{abbas2016personalized}. It is a cloud computing based risk assessment prediction system. In a system anticipated by Veeresh Patel et al, genomics and clinical data from patients EHR will be utilised to predict the cancer risk \cite{patel2015big}. Physical therapy-as-a-service (PTaaS) application is a model designed to connect the sensors available in the patient’s home and the therapist office computer\cite{calyam2016synchronous}. 

The above said healthcare system are using the well defined bussiness rules and vocabularies. But, health IT systems need different type of data management architecture to address its particular challenges. Particularly, healthcare IT needs a late binding model, which is flexible, time efficient, scalable, and adaptable\cite{barlow2017comparing}. 

Blockchain data structures utilises the data lakes to support the data from variety of sources such as patients’ mobile applications, wearable sensors, EMR’s, documents and images\cite{linn2016blockchain}. Blockchain data structures utilises the data lakes to support the data from variety of sources such as patients’ mobile applications, wearable sensors, EMR’s, documents and images.

\section{Conclusion}
The traditional data warehouse (DW) technique is no longer suitable for healthcare analytics due to its schema on write nature and inability to handle data silos. In our study, an effective data lake system was proposed for the personalized healthcare data recommendations. The data lake is a flat, schema-on-read data architecture. We conducted experiments to evaluate and compare the proposed data lake architecture and DW on the dataset obtained from UCI repository. Our data lake architecture outperformed DW significantly in terms of data ingestion time: The time for loading and storing data in data lake architecture is nearly 50\% less than that in DW. Moreover, the data lake architecture has higher precision in clustering than DW, mainly because of its ability to connect with more data sources. Briefly speaking, it was demonstrated that the data lake architecture can be an effective alternative to the existing health IT infrastructure. The proposed system can ingest all data in the unstructured, semi-structured, and structured formats. It can store data with low cost, taking advantages of HDFS. Hence, the data lake-based healthcare recommendation system will address the drawbacks of the traditional data architectures and provide additional capabilities for the future caliber of the data reusability.

\bibliographystyle{splncs}
\bibliography{ref}

\end{document}